\documentclass[aps,twocolumn,floatfix,superscriptaddress,a4paper,twoside]{revtex4}
\usepackage{amsmath,fixmath}
\usepackage{graphicx}
\usepackage{amssymb}
\usepackage{bm}
\usepackage[utf8]{inputenc}
\usepackage[english]{babel}

\newcommand{\V}[1]{\boldsymbol{#1}}
\newcommand{\be}[0]{\begin{equation}}
\newcommand{\ee}[0]{\end{equation}}
\newcommand{\mean}[1]{\left\langle {#1} \right\rangle}
\newcommand{\paren}[1]{\left( {#1} \right)}
\newcommand{\caja}[1]{\left[  {#1} \right]}
\newcommand{\lazo}[1]{\left\{  {#1} \right\}}

\begin{document}
\title{Spin Glasses on the Hypercube}
\author{L.A.~Fern\'{a}ndez} \affiliation{Departamento
  de F\'\i{}sica Te\'orica I, Universidad
  Complutense, 28040 Madrid, Spain.} 
  \affiliation{Instituto de Biocomputaci\'on y
  F\'{\i}sica de Sistemas Complejos (BIFI), Zaragoza, Spain.}
\author{V.~Martin-Mayor} \affiliation{Departamento de F\'\i{}sica
  Te\'orica I, Universidad Complutense, 28040 Madrid, Spain.} 
\affiliation{Instituto de Biocomputaci\'on y
  F\'{\i}sica de Sistemas Complejos (BIFI), Zaragoza, Spain.}
\author{G.~Parisi} \affiliation{Dipartimento di Fisica, INFM-CNR (SMC), Universit\`a di Roma ``La Sapienza'', 00185 Roma, Italy.}
\author{B.~Seoane} \affiliation{Departamento de F\'\i{}sica
  Te\'orica I, Universidad Complutense, 28040 Madrid, Spain.} 
\affiliation{Instituto de Biocomputaci\'on y
  F\'{\i}sica de Sistemas Complejos (BIFI), Zaragoza, Spain.}
\date{\today}
\begin{abstract}
We present a mean field model for spin glasses with a natural notion of
distance built in, namely, the Edwards-Anderson model on the diluted
$D$-dimensional unit hypercube in the limit of large $D$. We show that finite
$D$ effects are strongly dependent on the connectivity, being much smaller for
a fixed coordination number. We solve the non trivial problem of generating
these lattices. Afterwards, we numerically study the nonequilibrium dynamics
of the mean field spin glass. Our three main findings are: (i) the dynamics is
ruled by an infinite number of time-sectors, (ii) the aging dynamics
consists on the growth of coherent domains with a non vanishing surface-volume
ratio, and (iii) the propagator in Fourier space follows the $p^4$ law. We study as well finite $D$ effects in the nonequilibrium dynamics, finding that a naive finite size scaling ansatz works surprisingly well.
\end{abstract}
\maketitle 

\section{Introduction}
Spin Glasses (SG) are highly disordered magnetic systems
\cite{MYDOSH}. Rather than by their practical usefulness, SG are often
studied as a paradigmatic example of a complex system. Indeed, they
display an extremely slow dynamics on a complex free-energy landscape,
with many degenerate states. In addition, SG are a convenient
experimental model for glassy behavior, due to the comparatively fast
microscopic spin dynamics, as compared, for instance, with supercooled
liquids. In fact, nowadays, SG are considered as a playground to learn
about general glassy behavior, minimization problems in Computing
Science, biology, financial markets, etc.

Maybe the most conspicuous feature of SG is \textit{Aging}: they never
reach thermal equilibrium in experimental times. Here we will only
consider the simplest possible experimental protocol, the temperature
quench (see~\cite{EXPMEMORIA} for very interesting, more sophisticated
experimental procedures): the sample is cooled below the critical
temperature, $T_\mathrm{c}$, and it is let to relax for a time
$t_\mathrm{w}$ at the working temperature $T$. Its properties are
studied at a later time $t+t_\mathrm{w}$. It turns out that, if a
magnetic field was applied from the temperature quenched until
$t_\mathrm{w}$, when it is switched off, the thermoremanent
magnetization $M(t,t_\mathrm{w})$ decays with $t/t_\mathrm{w}$ (at
least for $10^{-3}<t/t_\mathrm{w}<10$ and
$50\,\text{s}<t_\mathrm{w}<10^4\,\text{s}$ \cite{RODRIGUEZ}). This
lack of a characteristic time beyond $t_\mathrm{w}$, the glassy system
age, is known as \textit{Full Aging}. We now know that Full Aging is
an effective description of the dynamics, no longer valid for
$t/t_\mathrm{w}\sim 10^{4}$ and $t_\mathrm{w}\sim 10\,\text{s}$
\cite{KENNING}.

Nowadays, we know that the slow dynamics in SG is originated by a
thermodynamic phase transition at $T_\mathrm{c}$
\cite{EXPERIMENTOTC}. Below $T_\mathrm{c}$, the spins associate in
coherent domains, whose size, $\xi(t_\mathrm{w})$, grows with
time. The lower the temperature, the slower the growth of
$\xi(t_\mathrm{w})$ is [experimentally,
  $\xi(t_\mathrm{w}\!=\!100\,\text{s};T\!=\!0.9T_\mathrm{c})\!\sim
  \!100$ atomic spacings].

There is a lively theoretical debate on the properties of the low temperature phase.  Surprisingly enough, this
controversy on the equilibrium properties of a non-accessible (in human
time scales) spin glass phase is relevant to {\em nonequilibrium}
experiments~\cite{FRANZ}.

Mainly, there are two competing theories, the \textit{droplets}
\cite{DROPLET}, and the replica symmetry breaking (RSB) one \cite{RSB}.
According to the \textit{droplets} picture, the SG phase would be
ferromagnetic-like, with a complicated spin texture, but essentially with only
two equilibrium states. On the other hand, the RSB theory predicts an infinite
number of degenerated states with an ultrametric organization. For both
theories, \textit{Aging} would be a coarsening process, in the sense that
coherent domains of low temperature phase grow with time. The two theories
disagree in their predictions for these domains properties. According to
\textit{droplets}, the domains would be compact objects, with a surface-volume
ratio that vanishes in the high $\xi(t_\mathrm{w})$ limit
\cite{SUPERUNIVERSALIDAD}. The SG order parameter is non zero inside of each
domain. On the contrary, the RSB theory expects non-compact domains with a
surface-volume ratio constant for large $\xi(t_\mathrm{w})$. Furthermore, in a
RSB system, the SG order parameter vanishes inside those domains. In recent
times, a somehow intermediate theory (TNT), has been proposed \cite{TNT}, but,
to our knowledge, no detailed dynamic predictions have been provided.

The RSB theory is based on the Mean Field approximation (MF) which,
unlike in the ferromagnetic case, is highly non trivial. Indeed, after
30 years of study, \textit{Aging} is not yet fully quantitatively
understood, not even within MF approximations. Therefore, non
perturbative tools, such as Monte Carlo (MC) calculations, appear as
an appealing alternative. Furthermore, MC calculations are called for
even at the MF level. Hence, one is interested in mean field models,
that is to say models where the MF approximation becomes exact in the
thermodynamic limit (TL). The standard MF model, (the
Sherrington-Kirkpatrick model, see \cite{SK} and
Sect.~\ref{sec:modelos}), has a number of disadvantages. It lacks a
natural notion of distance (hence one cannot discuss a coherence
length $\xi(t_\mathrm{w})$), or coordination number. Furthermore, its
numerical simulation is computationally heavier. In fact, recent
advances on the analytical study of spin glasses on Bethe
lattices~\cite{MEZARD-PARISI} has shifted the attention to these far
more numerically tractable models which share with experimental
systems the notion of a coordination number.

Here we wish to present a new MF model for spin-glasses: the
spin-glass on a $D$-dimensional hypercube with {\em fixed}
connectivity~\cite{MARINARI95}.  In the thermodynamic limit (which coincides with the
large $D$ limit for this model), Bethe approximation becomes exact.  As a
consequence, the statics is of Bethe-lattice type and can be computed.
A nice feature of this new model, is that it has a natural definition of distance,
which allow us to study spatial correlations within MF approximation.
In other words: this MF model is more similar to a real
$D\!=\!3$ system than any other studied before or, at least, than those
considered previously, since the space-time correlation functions can be
studied.

 The structure of this paper will be the following. In Sect. \ref{sec:modelos}
 we will describe the model and compare it with other MF models. In
 particular, in Sect. \ref{sec:fixed-con} we will address the problem of
 fixing the connectivity in a diluted hypercube. In Sect. \ref{sec:MN} we will
 briefly explain the numerical methods we have used and in Sect. \ref{sec:obs}
 we will introduce the observables measured during the simulations. In
 Sect. \ref{sec:res_eq} and \ref{sec:res_feq}, the numerical results will be
 presented, both in equilibrium (as a test of the model) and nonequilibrium,
 respectively. In Sect. \ref{sec:TF} we will discuss finite size effects. The
 analysis will reveal the $p^4$ propagator~\cite{DEDOMINICIS}, for the first time in a
 numerical investigation. Our conclusions will be presented in
 Sect. \ref{sec:concl}. Finally, we include extended discussions in two
 appendices.

\section{Models}\label{sec:modelos}
The standard model in SG is the Edwards-Anderson (EA) model. We will work with two kinds of degrees of freedom: dynamical and
quenched. The dynamical ones correspond to the spins, $\sigma_{i}$,
with $i\!=\!1,2,\ldots N$. We will consider them as Ising variables
$\pm1$. The non dynamical (or quenched) represent the material
impurities. We will consider here two types of them: the connectivity matrix, $n_{ik}\!=\!n_{ki}\!=\!1,0$ ($n_{ik}\!=\!1$ as long as spins
$i$ and $k$ interact), and the coupling constants,
$J_{ik}\!=\!J_{ki}$, which shall take only two opposite values (in
general with certain exceptions that will be discussed in the next
paragraph, we will consider $J_{ik}\!=\!\pm 1$, which defines our
energy scale). The interaction energy is \be\label{eq:H}
\mathcal{H}=-\sum_{i<k}J_{ik}n_{ik}\sigma_{i}\sigma_{k}\,.\ee Since
the impurity diffusion time is huge compared to the \textit{spin-flip}
(picosecond), we will always work within the so-called quenched
approximation: spins cannot have any kind of influence over the
material impurities. Then, both the set of coupling constants in the
Hamiltonian and its associated Gibbs free energy, will be considered
random variables. Therefore, in order to rationalize the experiments,
the useful free energy will be an average over the disorder.  We will
refer to each assignment of $\lazo{n_{ik},J_{ik}}$ as a
\textit{sample}. Its probability distribution defines the actual EA
model. The average over samples will be represented as
$\overline{(\ldots)}$.

Only a few exact results are known for the Hamiltonian \eqref{eq:H},
and all them were obtained within the MF approximation
\cite{RSB}. This \textit{approximation} becomes exact in a weak infinite-range
interaction model (in a ferromagnet $n_{ik}\!=\!1$ and
$J_{ik}\!=\!1/N$ for each couple $i,k$). In SG it is usually
represented as $n_{ik}\!=\!1$ for every couple $i,k$ and, because of
the random ferromagnetic and antiferromagnetic interaction character,
$\{J_{ik}\}$ are independent gaussian random variables, with zero mean
and variance $1/N$ \cite{SK}.

Computer simulations of long-range models are extremely hard, because
the energy evaluation for a system of $N$ spins requires $N^2$
operations. The situation has improved since the discovery that EA
models on Bethe lattices (not to be confused with Bethe trees) undergo
Replica Symmetry Breaking at $T_\mathrm{c}$~\cite{MEZARD-PARISI}.

A very popular realization of a Bethe-lattice spin glass is the EA model on a
Poisson graph. A simple way of drawing one of these graphs consists on
connecting ($n_{ik}\!=\!1$) each possible pair of spins, $i,k$, (there are
$N(N-1)/2$ possible couples), with probability $z/(N-1)$.  Thus, the number of
neighbors of spin $i$, its coordination number $n_i$ follows in the large-$N$
limit a Poisson distribution function with average $z$ (the connectivity). We
will consider $z\!=\!6$ to mimic a three dimensional system. This kind of
graphs are locally cycle-less: the mean shortest length among all the closed
loops that passes through a given point is $O(\log N)$, i.e. the system is
still locally tree-like. Computationally convenient as they are, Poisson
graphs still lack a natural notion of distance.

A simple alternative consists on formulating the model on a $D$-dimensional
unit hypercube. Thus, the spins are located in each of the hypercube vertices
(then, $N\!=\!2^D$) and the bonds lie on the edges. Therefore, each spin can
be connected with, at most, $D\!=\!\log_2 N$ spins.  By analogy with the
Poissonian graph, we consider that a link is active (i.e. $n_{ik}=1$) over
each edge with probability $z/D$. We call this model \textit{random
  connectivity hypercube}. It is easy to prove that it is locally tree-like as
well: the density of closed loops of length $l$ decays, at least, with
$D^{-2}$ (i.e. with the squared logarithm of $N$, as it also happens in the
Poisson graph). Incidentally, one could consider as well a non diluted hypercube,
but this would have two shortcomings: the connection with three dimensional
systems would get lost, and the numerical simulation would become
computationally heavy for large $D$.

Note that, at variance with other infinite-dimensional graphs, the
hypercube has at least two natural notions of distance: Euclidean
metrics and the postman metrics\footnote{The distance between two
  points, $\V{x}$ and $\V{y}$, is given by the minimum number of
  edges, either occupied or not, that must be covered when joining
  $\V{x}$ and $\V{y}$.}. The two distances are essentially equivalent,
since the Euclidean distance between two sites in the hypercube is
merely the square root of the postman distance. In the following we
shall use the postman metrics, which has some amusing consequences. For instance, our correlation-length will be the {\em square} of the
Euclidean one, thus yielding a critical exponent $\nu=1$, doubling the
expected $\nu_\text{MF}=1/2$. Of course, if we use the Euclidean metric we recover the usual exponent $\nu=1/2$.

However, it turns out that the random connectivity hypercube suffers a
major disadvantage. The inverse of the critical temperature in a
ferromagnet \cite{PARISI} or in a SG \cite{KC} can be computed within
the Bethe approximation:\be\label{eq:KC}
K_\mathrm{c}^\mathrm{FM}=\mathrm{atanh}\frac{1}{\mean{n}_1-1}\ ,\ K_\mathrm{c}^\mathrm{SG}=\mathrm{atanh}\frac{1}{\sqrt{\mean{n}_1-1}}\,.
\ee In this expression $\mean{n}_1$ is a conditional expectation value
for $n$, the coordination number of a given site in the graph.  This
conditional expectation value is computed knowing for sure that our
site is connected to another {\em specific} site (this is different
from the average number of neighbors of a site that has at least one
neighbor!). A simple calculation shows that
$\mean{n}_1\!=\!1+z-\frac{z}{D}$ in the random connectivity
model. Since $D=\log_2 N$, we must expect huge finite size corrections
($O(1/\log N)$) at the critical point. Note that this problem is far
less dramatic for a Poisson graph where
$\mean{n}_1\!=\!1+z-\frac{z}{N-1}$.

The cure seems rather obvious: place the occupied links in the
hypercube in such a way that $n=z$ (here, $z=6$). Unfortunately,
drawing these graphs poses a non trivial problem in Computer Science
\cite{RICCI}. Our solution to this problem is discussed in the next
paragraph.

\subsection{The fixed connectivity hypercube}\label{sec:fixed-con}

We have not found any systematic way of activating links in the hypercube
that respects the fixed connectivity condition. Thus, we have adopted
an operational approach: the distribution of bonds is obtained by
means of a dynamic MC. 
We must define a MC procedure that generates a set of graphs that remains
invariant under all symmetry transformations of the hypercube group.

Specifically, we start with an initial condition in which all bonds
along the directions 1 to 6 are activated (of course, this procedure makes sense only for $D\!\ge\!6$). Clearly enough,
the initial condition verifies the constraint $n=6$. We shall modify
the bond distribution by means of movements that do not change $n$. We
perform what we called a ``plaquette'' transformation (a plaquette is
the shortest possible loop in the hypercube, of length 4).  We
randomly pick, with uniform probability, one hypercube plaquette. In
case this plaquette contains only two parallel active links
($n_{ik}=1$), these two links are deactivated at the same time that
the other two are activated. On the opposite case, nothing is
done \footnote{This movement keeps each vertex connectivity
  unaltered. Besides, a transformation and its opposite are equally
  probable. As a consequence the Detailed Balance Condition is
  satisfied with respect to the uniform measure on the ensemble of
  fixed connectivity graphs. An standard theorem \cite{LIBRO} ensures
  that the equilibrium state of this Markov chain is the uniform measure over
  the subset of fixed connectivity hypercubes reachable from the
  initial condition by means of plaquette transformations}. This
guarantees that the set of generated graphs is isotropic.

\begin{figure}
\begin{center}
 \includegraphics[angle=270,width=\columnwidth,trim=20 30 20 40]{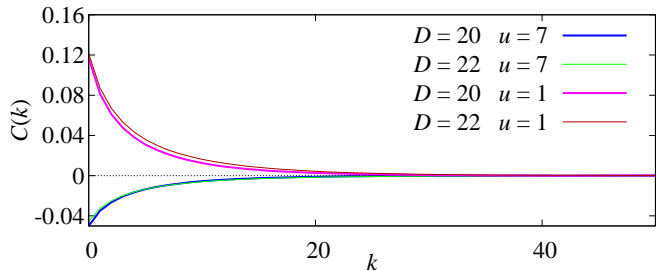}
\caption{ Generation algorithm of fixed
  connectivity graphs: for two system sizes ($D\!=\!20,22$) and two spatial
  directions ($u\!=\!1,7$), we represent the density of occupied edges as
  function of the MC time. As MC time goes on, the system recover the lost
  isotropy induced by the initial condition.}\label{fig:corrtimes}
 \end{center}
\end{figure}

In order to this procedure to be useful, the dynamic MC correlations
times must be short. In Fig. \ref{fig:corrtimes}, we show the MC
evolution of the system isotropy. We make $kN$ plaquette
transformations, and we control the density of occupied bonds in two
directions: the first direction (initially occupied in every vertex)
and the seventh direction (initially unoccupied). As we see, for two
different system sizes, we get short isotropization exponential times
(for $D\!=\!22$ we get $\tau_\mathrm{exp}\!\approx \! 4.7 N$). For
this reason, we assume that taking $k\!=\!100$ is long enough to
ensure that the configurations obtained are completely independent
from the initial condition.

At this point, the question arises of the completeness of the generated set of
graphs. We first note that our set contains proper subsets that are also
isotropic. However, finding them would require more involved algorithms which
will not pay in a reduction of statistical errors (as we will show below, most
of the sample dispersion is induced by the coupling matrix
$\lazo{J_{ik}}$). On the other hand, one could think that there are lacking
graphs in our algorithm for a simple reason. The plaquette transformation
cannot break loops: when we interchange neighboring links we can only either
join two different loops or split up a loop into two loops. Due to the
hypercube boundary conditions, in the initial configuration all sites belonged
to closed loops. This situation cannot be changed by plaquette
transformations. However, this objection does not resist a close
inspection. In fact, a non-closed lattice path formed by occupied links should
have an ending point with an {\em odd} coordination number, which violates the
constraint $n=z$ for any even $z$. Thus, all lattice paths compatible with our
fixed connectivity constraint, do form closed loops.  This
argument, as well as the numerical checks reported below, make us confident
that the set of generated graphs is general enough for our
purposes. Actually, we conjecture that our algorithm
  generates \textit{all} possible fixed connectivity graphs with $z$ even.

One may worry as well about the applicability of the Bethe approximation to
the fixed connectivity model, since all loops are closed. Actually, the
crucial point to apply the Bethe approximation is that the probability of
having a closed path of any {\em fixed} length should vanish in the large $D$
limit. This is easy to prove for the random connectivity model. In the fixed
connectivity case, one may argue as follows. Let us imagine a walk over the
closed path. On the very first step, the probability that the chosen link is
present is $z/D$, whereas in the following step the probability of finding the
link is $(z-1)/(D-1)$ in the limit of large $D$ (since one of the $z$ links
available at the present site was already used to get there).  This estimate
implicitly assumes that the occupancy of different links is statistically
independent. The independency approximately holds for large $D$ and becomes
exact in the $D\to\infty$ limit, where occupied links form a diluted set. At
this point, the estimate of the number of paths of any given fixed length in
the large $D$ limit can be performed as in the random-connectivity case. One
finds as well, in the fixed connectivity case, that the number of closed loops
per site of a given length decays at least as $O(1/D^2)$.

\begin{figure}
\begin{center}
 \includegraphics[angle=270,width=\columnwidth,trim=40 40 20 40]{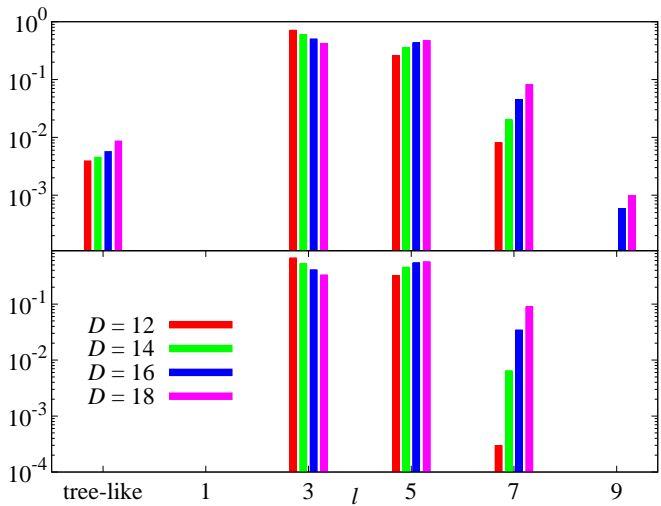}
\caption{For (\textbf{top}) random-$z$ graphs and (\textbf{bottom})
  Fixed-$z$ for plaquette transformations, the probability
  distribution function of the length of the second shortest path
  joining nearest-neighbors for different sets of graphs (mind the
  vertical axis is in logarithmic scale). Lines has been slightly
  displaced in order to help the
  visualization. }\label{fig:min_distance}
\end{center}
\end{figure}

In addition to the above considerations, we may numerically compute in
our graphs the length of the second shortest path that joins two {\em
  connected} nearest neighbors in the hypercube. In
Fig. \ref{fig:min_distance}, we compare the probabilities for the
length of such paths in the random (top) and fixed (bottom)
connectivity models, for different system sizes \footnote{We use a very simple algorithm to find the length of the second shortest path joining two neighboring spins $i_1,\ i_2$. We consider a truncated connectivity matrix, $\tilde{n}$, that coincides with the true one, $n$, but for the link $i_1-i_2$, which is deactivated: $\tilde n_{i_1,i_2}=\tilde n_{i_2,i_1}=0$. We take a starting vector $\V{v}^{(0)}$ with all its components set to zero but the component $i_1$ which is set to one. We iteratively multiply the vector by the truncated connectivity matrix, i.e. $\V{v}^{(t)}=\tilde n \V{v}^{(t-1)}$, until the $i_2$-th component is nonzero. The sought length is just the minimum value of $t$ that fulfills the stopping condition.}. In both cases, we
note that the maximum of the probability shifts to larger length as
$D$ grows. We note as well that, for fixed connectivity, no tree-like
graph arises\footnote{We say that a graph is a tree-graph if, once the
  link between two neighboring spins is removed, there is no way
  of joining them following any other path.}.

\begin{figure}
\begin{center}
 \includegraphics[angle=270,width=\columnwidth,trim=40 15 20 40]{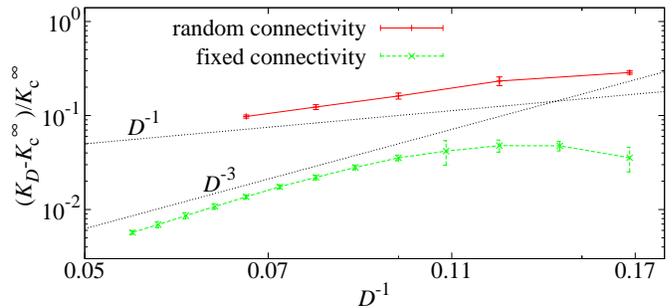}
      \caption{Comparison of finite volume effects in the critical point
  estimators $K_\mathrm{c}^D$ for ferromagnetic Ising model, both in the
  random \textbf{(red dots)} and fixed \textbf{(green crosses)}connectivity hypercubes. As a guide to the eye, we
  have included two different scalings with $D$. The estimator
  $K_\mathrm{c}^D$ corresponds to the average of the inverse
  temperatures at which the Binder cumulant, Eq. \eqref{eq:binder},
  reaches the values 1.2 and 2.4.}\label{fig:KD}
\end{center}
\end{figure}

A summary of our efforts is shown in Fig. \ref{fig:KD}, where we plot
the evolution of the critical point with $D$ for the ferromagnetic
Ising model, defined on hypercubes with both random and fixed
connectivity. As expected, the random connectivity model suffers very
important finite volume corrections which make it essentially useless
for numerical studies. The problem is solved using fixed connectivity
hypercubes instead, where the finite volume effects are only caused by
the residual presence of short closed loops.

\section{Numerical Methods}\label{sec:MN}

We have simulated the Hamiltonian \eqref{eq:H} using a Metropolis
algorithm \cite{LIBRO}. In addition, we use \textit{Multispin Coding}:
since spins are binary variables, we can simultaneously codify 64
systems in one single 64 bits word (all of them share the same connectivity
matrix, $n_{ik}$). With this common matrix, we find errors which are
$\!\sim$7 times smaller than those obtained with one single sample per
matrix. This should be compared with the factor 8 that we would obtain
in case no correlation were induced. Our program needs
$0.29\,\text{ns}$/\textit{spin-flip} in an Intel i7 at 2.93GHz (in
Ref. \cite{HASENBUSCH} they report
$\!\sim\!1.2\,\text{ns}$/\textit{spin-flip} on an Opteron at 2.0 GHz,
for the simulation of the $D=3$ EA model in the cubic lattice).

In a nonequilibrium dynamical study such as ours, one computes both one-time and
two-times quantities, see Sect. \ref{sec:obs}. The calculation of
two-times quantities implies the storage on disk of intermediate
configurations. Disk capacity turned out to be the main limiting factor
for the simulation. For this reason, we have worked in parallel with two
program versions: one valid for measuring quantities at one and two
times and another restricted to  the computation of  one-time quantities.

We have computed two-time quantities at temperature $T=0.7T_\text{c}$,
on systems with $D=16,18,20$ and $22$. The number of simulated samples
were $8\!\times\! 64$ samples for each system size (hence, for
self-averaging quantities the statistical quality of our data grow
with $D$).

Besides, since this new model requires intensive testing, we have
computed {\em equilibrium} one-time quantities at
$T/T_\text{c}=0.95,0.97,0.99,1,1.1,1.2,1.3$ and $1.4$. The system
sizes were again $D=16,18,20$ and $22$. The number of simulated
samples was $128\!\times\! 64$ samples per temperature (at
$T_\mathrm{c}$ we computed $256\!\times\! 64$ samples).

\section{Observables}\label{sec:obs}
The Hamiltonian \eqref{eq:H} has a global symmetry $\mathbb{Z}_2$ ($\sigma_{i}\rightarrow-\sigma_{i}$ for all $i$). Not as obvious is the gauge
symmetry induced by the average over couplings. In fact, choosing randomly a sign for each position, $\varepsilon_{i}\!=\!\pm 1$, the energy \eqref{eq:H} is
invariant under the transformation 
\be\label{eq:gi}\begin{array}{cc}
\sigma_{i}\rightarrow  \varepsilon_{i}\sigma_{i}\,,\ &
J_{ik}\rightarrow  \varepsilon_{i}\varepsilon_{k}J_{ik}\,.\\
   \end{array}
\ee
Now, since the transformed couplings $ \varepsilon_{i}\varepsilon_{k}J_{ik}$ are just as
probable as the original ones, we need to define observables that are
invariant under the gauge transformation \eqref{eq:gi}. With this aim we form gauge invariant fields from two systems at equal time, that evolve
independently with the same couplings,
$\{\sigma_{i}^{(1)},\sigma_{i}^{(2)}\}$ (real replicas) or, alternatively, from a single system considered at two
different times:
\be \begin{array}{c}q_{i}(t_\mathrm{w})=\sigma_{i}^{(1)}(t_\mathrm{w})\sigma_{i}^{(2)}(t_\mathrm{w})    \,\text{, }   \\c_{i}(t,t_\mathrm{w})=\sigma_{i}^{(1)}(t+t_\mathrm{w})\sigma_{i}^{(1)}(t_\mathrm{w})\,.\end{array}\ee

We can define three kinds of quantities with both fields.

\textbf{1. One-time-quantities.}
The order parameter
\be q(t_\mathrm{w})=\sum_{i}q_{i}(t_\mathrm{w})\,,\ee
vanishes in the nonequilibrium regime (the system is much bigger than the
coherence length, $\xi(t_\mathrm{w})$).
The non linear susceptibility is proportional to the SG susceptibility:
\be\chi_\mathrm{SG}(t_\mathrm{w})=N\overline{q^2(t_\mathrm{w})}\,,\ee
that grows with a power of $\xi(t_\mathrm{w})$.
The Binder parameter provide us with information about the fluctuations
\be\label{eq:binder} B(t_\mathrm{w})=\frac{\overline{q^4(t_\mathrm{w})}}{\overline{q^2(t_\mathrm{w})}^2}\,.\ee
In the Gaussian regime $B\!=\!3$. In a ferromagnetic phase, $B\!=\!1$.
In the SG phase, in equilibrium (that for finite volume corresponds to $t_\mathrm{w}\to\infty$), 
$B$ grows with the temperature from $B\!=\!1$ at $T=0$. The equilibrium paramagnetic
phase is in Gaussian regime.

\textbf{2. Two-time-quantities.}
The correlation spin function tells us about the memory kept by the system, at time $t+t_\mathrm{w}$, about the configuration at $t_\mathrm{w}$~\footnote{We storage configurations
  at times $2^n+2^m$, with $n,m$ integers. We calculate the correlation
  function for all $(t,t_\mathrm{w})$ power of 2, allowed by the simulation length.}:
\be
C(t,t_\mathrm{w})=\frac{1}{N}\overline{\sum_{i}c_{i}(t,t_\mathrm{w})}\,.\ee
The susceptibility is
$\chi(\omega\!=\!2\pi/t,t_\mathrm{w})\!\propto\![1-C(t,t_\mathrm{w})]/T$.
On the other hand, when $t_\mathrm{w}$ is fixed, $C(t,t_\mathrm{w})$ is just
the thermoremanent magnetization~\footnote{Using the gauge transformation
  \eqref{eq:gi}, it is possible to rewrite an ordered configuration (by an
  external magnetic field, for instance), as the spin configuration found at
  time $t_\mathrm{w}$ after a random start.}.

The link correlation function 
\be C_\mathrm{link}(t,t_\mathrm{w})=\frac{1}{DN}\overline{\sum_{ik}n_{ik}\ c_{i}(t,t_\mathrm{w})c_{k}(t,t_\mathrm{w})}\,,\ee
carries the information of the density of the interfaces between coherent
domains at $t_\mathrm{w}$, that at $t+t_\mathrm{w}$, have flipped. In case the
surface-volume ratio decayed with a negative power of $\xi(t_\mathrm{w})$
(\textit{droplets}), $C_\mathrm{link}$ would become $t$-independent~\cite{JANUSJSP}. On the
contrary, in a RSB system, $C_\mathrm{link}\!=\!a+b\,C^2$. Note that, for the Sherrington-Kirkpatrick model, one trivially finds $C_\mathrm{link}=C^2$, but the linear relation is not straight-forward in fixed connectivity mean field models.

\begin{figure}
\begin{center}
\includegraphics[angle=270,width=\columnwidth,trim=40 15 20 40]{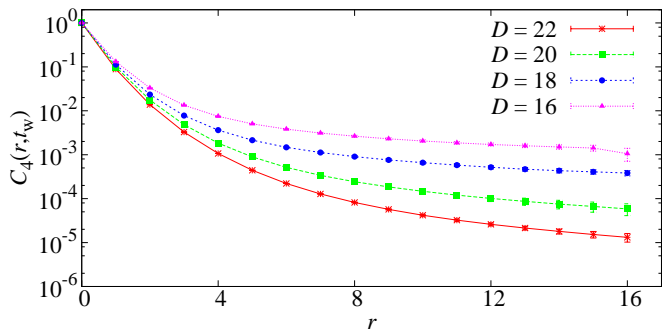}
\caption{$C_4(r,t_\mathrm{w})$, Eq. \eqref{eq:fcor_sinJ}, for
  $t_\mathrm{w}\!=\!2^8$ and different system sizes, $N\!=\!2^D$, at $T\!=\!0.7T_\mathrm{c}$.}\label{fig:CR}
\end{center}
\end{figure}

\textbf{3. Spatial correlation functions.} 
In the unit hypercube, the binary decomposition of the spin index
$i\!=\!1,2,\ldots,2^D$ can be equal to its Euclidean coordinates,
$\V{x}$. The spatial correlation function is 
\be c_4(\V{r},t_\mathrm{w})=\frac{1}{N}\overline{\sum_{\V{x}}q_{\V{x}}(t_\mathrm{w})q_{\V{x}+\V{r}}(t_\mathrm{w})}\,.\ee
We consider $r\!=\!|\V{r}|$ as the distance in the postman
metrics. It would look rather natural to average
all the $c_4(\V{r},t_\mathrm{w})$ over the $N_r\!=\!\binom{D}{r}$
displacements of length $r\!=\!|\V{r}|$:
\be\label{eq:fcor_sinJ} C_4(r,t_\mathrm{w})=\frac{1}{N_r}\sum_{
\V{r},
|\V{r}|=r
}c_4(\V{r},t_\mathrm{w})\,.\ee
However, see Fig. \ref{fig:CR}, $C_4(r,t_\mathrm{w})$ does not
present a limiting behavior with $D$ for a given $t_\mathrm{w}$.

\begin{figure}
\begin{center}
\includegraphics[angle=270,width=\columnwidth,trim=50 15 20 40]{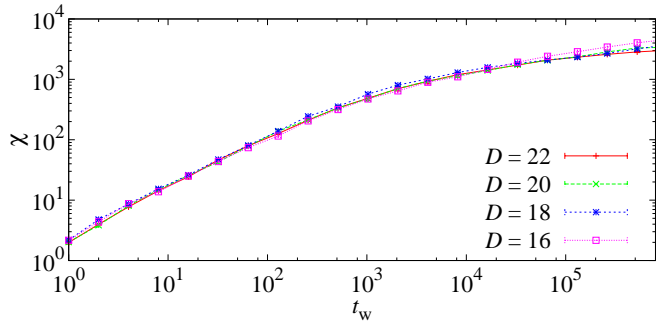}
\caption{SG susceptibility a $T\!=\!0.7T_\mathrm{c}$ as function of $t_\mathrm{w}$ 
for different system sizes, $N\!=\!2^D$.}\label{fig:chi_07}
\end{center}
\end{figure}

We can get a clue by looking at $\chi_\mathrm{SG}(t_\mathrm{w})$,
Fig. \ref{fig:chi_07}, which does reach a thermodynamic limit. Since
$\chi_\mathrm{SG}(t_\mathrm{w})$ is nothing but the integral of
$C_4(r,t_\mathrm{w})$ with the Jacobian $\binom{D}{r}$, it seems reasonable to
define the spatial correlation function instead:
\be\label{eq:fcor_conJ} \hat {C_4 }(r,t_\mathrm{w})=\sum_{
\V{r},
|\V{r}|=r
}c_4(\V{r},t_\mathrm{w})\,.\ee

\begin{figure}
\begin{center}
\includegraphics[angle=270,width=\columnwidth,trim=60 15 20 40]{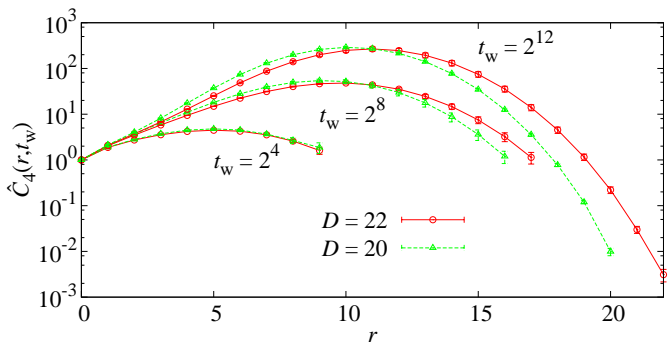}
\caption{$\hat {C_4 }(r,t_\mathrm{w})$, Eq. \eqref{eq:fcor_conJ},  for
  $D\!=\!10$ and 22 for $t_\mathrm{w}\!=\!2^4,\ 2^8$ and $2^{12}$ at
  $T\!=\!0.7T_\mathrm{c}$. This has to be compared with the behavior of $C_4(r,t_\mathrm{w})$, Fig. \ref{fig:CR}. }\label{fig:TCR}
\end{center}
\end{figure}

We can see that $\hat {C_4 }(r,t_\mathrm{w})$ does reach the high-$D$
limit, Figure \ref{fig:TCR}, at least for short
$t_\mathrm{w}$. Besides, in the paramagnetic phase, it is possible to compute analytically $\hat {C_4 }(r,t_\mathrm{w})$, see Appendix \ref{ap:HTE}, taking first the limit $t_\mathrm{w}\to\infty$ and making afterwards $D\to\infty$. The resulting correlation function, which is only valid in the paramagnetic phase, is a simple exponential. Hence, both the equilibrium and the nonequilibrium computations, suggest that one should focus on $\hat {C_4 }$ rather than on $C_4$.

We note in Fig. \ref{fig:TCR}, that in the SG phase, $\hat {C_4 }$ is non monotonically decreasing with $r$, but rather presents a maximum. This maximum moves to bigger $r$ with
$t_\mathrm{w}$, then, the system has a characteristic length that increases
with time.
Thus, we can estimate the coherence length, by means of the integral estimator
$\xi_{0,1}(t_\mathrm{w})$:
\be\label{eq:xi} \xi_{0,1}(t_\mathrm{w})=\frac{\int_0^\infty
  \mathrm{d}r\ r\ \hat {C_4 }(r,t_\mathrm{w})}{\int_0^\infty \mathrm{d}r\ \hat
  {C_4 }(r,t_\mathrm{w})}\,.\ee
A major advantage of $\xi_{0,1}$ over more heuristic definitions of the coherence length, is that it is computed from self-averaging quantities (see details in~\cite{PRLJANUS,JANUSJSP}, we note that, in this work, we have not tried to estimate the contribution to the integrals by the noise-induced long distance cutoff).

The existence of such a characteristic length is the main advantage of the hypercube model over other MF models.

\section{Equilibrium Results}\label{sec:res_eq}
Since the present work is the first study ever made of a EA model on a fixed connectivity
hypercube it is necessary to make a few consistency
checks. Equilibrium results are most convenient in this respect, since we have analytical computations (valid only for the large $D$ limit) to compare with.

We will briefly study the spatial correlations in the
paramagnetic phase. In addition, we will check, by approaching to
$T_\mathrm{c}$ from the SG phase, that the SG transition does lie on the
predicted $T_\mathrm{c}$, Eq. \eqref{eq:KC}.

\subsection{Paramagnetic Phase}
Our very first check will be the comparison between the Monte Carlo estimate of the SG susceptibility (at finite $D$) with the analytical computation for infinite $D$:
\be\label{eq:chi_infty}\chi(T)=1+\frac{z\tanh^2T^{-1}}{1-(z-1)\tanh^2T^{-1}},\ee
see Appendix \ref{ap:HTE}. Our results are presented in Table \ref{tab:chi}. We see that finite size effects increase while approaching $T_\mathrm{c}$. For our larger system, $D=22$, the susceptibility significantly deviates from the asymptotic result only in the range $T_\mathrm{c}<T<1.1T_\mathrm{c}$.

\begin{table}
\begin{center}
\begin{tabular*}{\columnwidth}{@{\extracolsep{\fill}}cccc}\hline
$T$&$\chi(T)_{D=\infty}$&$\chi(T)_{D=20}$&$\chi(T)_{D=22}$\\\hline
$1.4T_\mathrm{c}$&2.4497$\ldots$&2.41(3)&2.44(3)\\
$1.3T_\mathrm{c}$&3.0176$\ldots$&2.98(4)&2.98(4)\\
$1.2T_\mathrm{c}$&4.1650$\ldots$&4.08(6)&4.10(7)\\
$1.1T_\mathrm{c}$&7.6344$\ldots$&7.11(13)&7.43(11)\\
$T_\mathrm{c}$&$\infty$&26(2)&98(7)\\\hline
\end{tabular*}
\end{center}
\caption{Comparison between the SG susceptibility in large $D$ limit for the paramagnetic phase, Eq. \eqref{eq:chi_infty}, and numerical results for $D=20,22$.}\label{tab:chi}
\end{table}

\begin{figure}
\begin{center}
\includegraphics[angle=270,width=\columnwidth,trim=40 15 20 40]{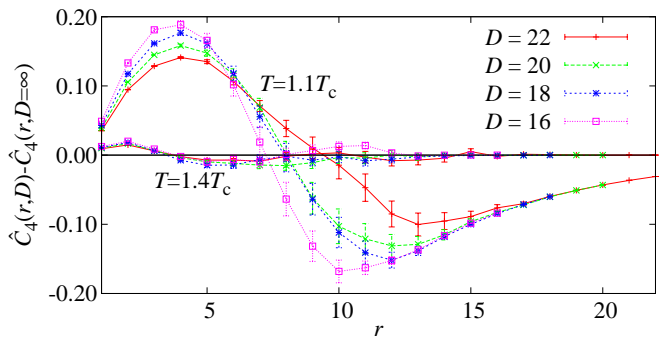}
\caption{Difference between the numerical and analytical spacial correlation
  function for different system sizes at two temperatures $T=1.1T_\mathrm{c}$
  and $T=1.4T_\mathrm{c}$.}\label{fig:CRpara}
\end{center}
\end{figure}

After the fast convergence to the large $D$ limit observed in the SG
susceptibility, the results for $\hat C_4$ are a little bit disappointing. In
Fig. \ref{fig:CRpara} we display $\hat C_4(r,D)-\hat C_4(r,\infty)$ as a
function of $r$.
We can see that finite size effects become more
important once one approaches $T_\mathrm{c}$.
Besides, finite $D$ corrections as a function of $r$ oscillate between positive and negative values. This is not surprising: the finite $D$ corrections to the susceptibility, which are very small, are just the integral under these curves.
More quantitatively, we see in Table \ref{tab:CRparaD} that the corrections with $D$ for $r=1,2$ are $O(D^{-1})$. Indeed, the path counting arguments in Appendix \ref{ap:HTE} are plagued by corrections of $O(D^{-1})$.

\begin{table}
\begin{center}
\begin{tabular*}{\columnwidth}{@{\extracolsep{\fill}}ccc|cc}\hline
&\multicolumn{2}{c|}{$r=1$}&\multicolumn{2}{c}{$r=2$}\\
$D$&$T=1.1T_\mathrm c$&$T=1.4T_\mathrm c$&$T=1.1T_\mathrm c$&$T=1.4T_\mathrm c$\\\hline
16&0.783(6)&0.198(5)&2.130(18)&0.320(12)\\
18&0.779(4)&0.201(3)&2.115(11)&0.327(7)\\
20&0.784(2)&0.202(2)&2.109(6)&0.332(4)\\ 
22&0.7776(12)&0.2006(9)&2.083(4)&0.324(2)\\\hline
\end{tabular*}
\end{center}
\caption{$D$ times the difference between $\hat C_4(r)$, for finite $D$ and infinite $D$, as computed for $r=1,2$. The absence of any $D$ evolution evidences finite-$D$ corrections of order $1/D$.}\label{tab:CRparaD}
\end{table}

\subsection{SG phase}
\begin{figure}
\begin{center}
\includegraphics[angle=270,width=\columnwidth,trim=40 30 20 40]{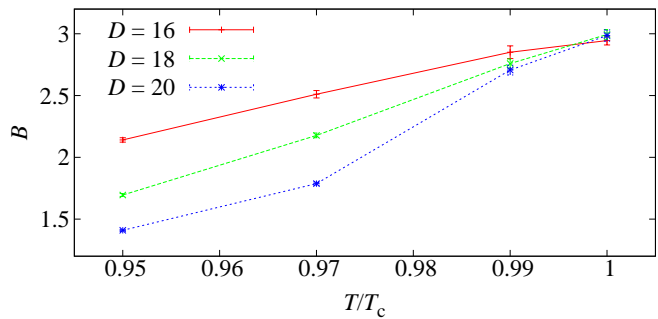}
\caption{Equilibrium values of the Binder cumulant, Eq. \eqref{eq:binder}, for
  several system sizes, as function of the temperature in units of the exact
  asymptotic value of $T_\mathrm{c}$, Eq. \eqref{eq:KC}, in the SG phase.}\label{fig:binder}
\end{center}
\end{figure}

In the SG phase, our test has been restricted to a check of Eq. \eqref{eq:KC}, that predicts a SG phase transition for the high-$D$
limit. With this aim, we compute the Binder cumulant, $B(T)$, nearby $T_\mathrm{c}$. For all $T<T_\mathrm{c}$, we expect $B(T)<3$ for large enough $D$. As we show in Fig. \ref{fig:binder}, $B(T)$ decreases with $T$ and shows sizeable finite size effects. In fact, at $T=0.99T_\mathrm{c}$, we need to simulate lattices as large as $D=20$ to find values below 3. Right at $T_\mathrm{c}$, the Gaussian value $B(T)=3$ is found for all the simulated sizes.

\section{Nonequilibrium Results}\label{sec:res_feq}
In this section we will address the main features of the nonequilibrium dynamics obtained in our largest system, $D=22$. The issue of finite $D$ corrections will be postponed to Sect. \ref{sec:TF}.
\subsection{The structure of isothermal aging}
The picture of isothermal aging dynamics in MF models of SG behavior was largely drawn in~\cite{CUGLIANDOLO} (see also \cite{YOUNG}). The dynamics is ruled by an infinite number of \textit{time-sectors}:
\be\label{eq:timesectors}
C(t,t_\mathrm{w})=\sum_{i}f_i\paren{h_i(t_\mathrm w)/h_i(t+t_\mathrm w)}.
\ee
The scaling functions $f_i$ are positive, monotonically decreasing and normalized, i.e. $1=\sum_i f_i(1)$. The unspecified functions $h_i$ are such that, in the large $t_\mathrm w$ limit, 
$h_i(t_\mathrm w)/h_i(t+t_\mathrm w)$ is 1 if $t\ll t_\mathrm w^{\mu_i}$, while it tends to zero if $t\gg t_\mathrm w^{\mu_i}$. In other words, the decay of $C$ between values $C_i$ and $C_{i+1}$ is ruled by the scaling function $f_i$ and takes place in the \textit{time-sector} $t\sim t_\mathrm w^{\mu_i}$. 

This picture is radically different to the \textit{Full Aging} often found both in experiments and in 3D simulations. A full aging dynamics is ruled only by two sectors of time, $\mu_i=0,1$. Nevertheless, recent experimental studies~\cite{KENNING} show that full aging is no longer fulfilled for $t\!\gg\!t_\mathrm{w}$. Probably more time-sectors must be considered to rationalize these experiments.

However, Eq. \eqref{eq:timesectors} is probably an oversimplification, since the spectrum of exponents $\mu_i$ might be continuous. An explicit realization of this idea was found in the critical dynamics of the trap model~\cite{BOUCHAUD}, where the correlation function behaves for large $t_\mathrm w$ as
\be\label{eq:ultrametric1}C(t,t_\mathrm{w})=f\paren{\alpha(t,t_\mathrm{w})}\,,\quad \alpha(t,t_\mathrm{w})=\log t/\log t_\mathrm{w}\,.\ee
Again, the scaling function $f$ is positive and monotonically decreasing. Clearly enough, in the limit of large $t_\mathrm w$ and for any positive exponent $\mu$, if $t=At_\mathrm w^\mu$, the correlation function takes a value that depends only on $\mu$, no matter the value of the amplitude $A$.

\begin{figure}
\begin{center}
\includegraphics[angle=270,width=\columnwidth,trim=40 25 20 40]{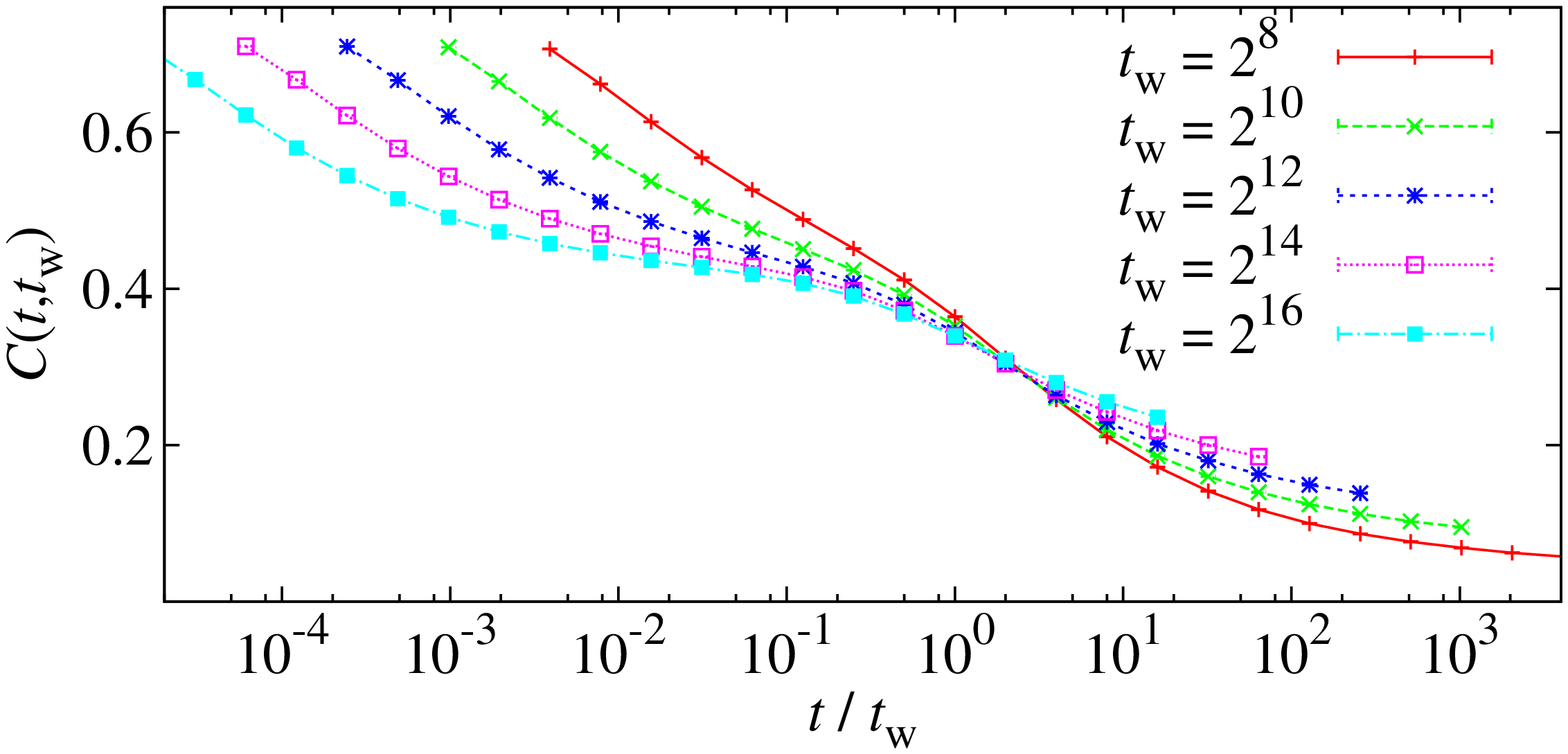}
\caption{$C(t,t_\mathrm{w})$ over $ t/ t_\mathrm{w}$ for $D\!=\!22$ and $T\!=\!0.7T_\mathrm{c}$.}\label{fig:fullaging}
\end{center}
\end{figure}

\begin{figure}
\begin{center}
\includegraphics[angle=270,width=\columnwidth,trim=40 25 20 40]{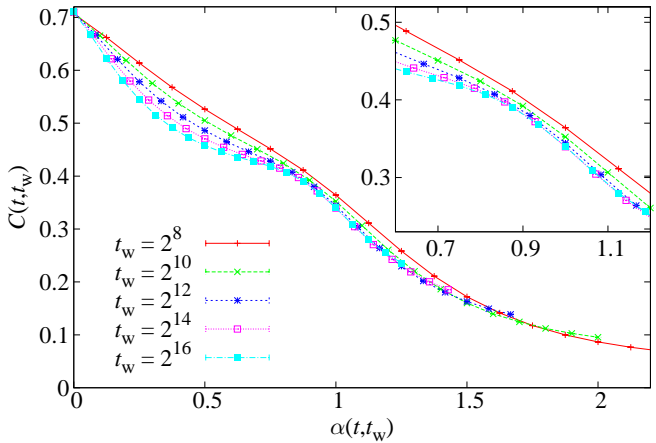}
\caption{Same data of Fig. \ref{fig:fullaging}, as a function of $\alpha(t,t_w)$, defined in Eq. \eqref{eq:ultrametric1}. The window is a zoomed image of the central region.}\label{fig:ultrametric1}
\end{center}
\end{figure}

As expected, $C(t,t_\mathrm{w})$ is clearly not a function of $t/t_\mathrm{w}$ in our model, see Fig. \ref{fig:fullaging}. On the contrary, data seem to tend to a constant value when $t_\mathrm w\to \infty$ in any finite range of the variable $t/t_\mathrm{w}$. This is precisely what one would expect in a time-sectors scheme.
On the other hand, if we try (without any supporting argument) the Bertin-Bouchaud scaling, Eq. \eqref{eq:ultrametric1}, see Fig. \ref{fig:ultrametric1}, the data collapse is surprisingly good. Therefore, the nonequilibrium dynamics in the SG phase seems ruled by a, not only infinite but continuous, spectrum of time-sectors.

We note \textit{en passant} that the scaling \eqref{eq:ultrametric1} is ultrametric only if the scaling function vanishes for all $\alpha(t,t_w)>1$, for details see Appendix \ref{ap:Ultrametricity}. In fact, dynamic ultrametricity is a geometric property~\cite{CUGLIANDOLO} that states that for all triplet of times $t_1\gg t_2\gg t_3$, one has in the limit $t_3\to\infty$:
\be C(t_1-t_3,t_3)=\min\left\{C(t_1-t_2,t_2),C(t_2-t_3,t_3)\right\}.\ee
Finding dynamical ultrametricity in concrete models has been rather elusive up to now. An outstanding example is the critical trap model~\cite{BOUCHAUD}, where $f(\alpha>1)=0$. It is amusing that the trap model is \textit{not} ultrametric from the point of view of the equilibrium states~\cite{MEZARD-PARISI-VIRASORO}. Thus, the casual connections between static and dynamic ultrametricity are unclear to us. At any rate, since our scaling function in Fig. \ref{fig:ultrametric1} does not show any tendency to vanish for $\alpha(t,t_w)>1$, we do not find compelling evidences for dynamic ultrametricity in this model.

We have also looked directly to the plots of $C(t_1-t_2,t_2)$ versus
$C(t_2-t_3,t_3)$ (see Appendix \ref{ap:Ultrametricity}) and we have not found convincing indications for the onset
of dynamical ultrametricity. In this respect, it is worth to recall similarly
inconclusive numerical investigations of the Sherrington-Kirkpatrick model~\cite{FALLOSSK}. There are two possible conclusions:
\begin{enumerate}
\item the model does not satisfy dynamical ultrametricity in spite of the
fact that it satisfies (according to the standard wisdom) static
ultrametricity.
\item Dynamical ultrametricity  holds but its onset is terrible slow.
\end{enumerate}
Both conclusions imply that it is rather difficult to use the dynamic
experimental data (or any kind of data) to get conclusions on static
ultrametricy. Of course it would be crucial to check if static
ultrametricity is satisfied in this model, but this is beyond the
scope of this paper.
\subsection{Aging in $C_\text{link}$}
\begin{figure}
\begin{center}
\includegraphics[angle=270,width=\columnwidth,trim=40 25 20 40]{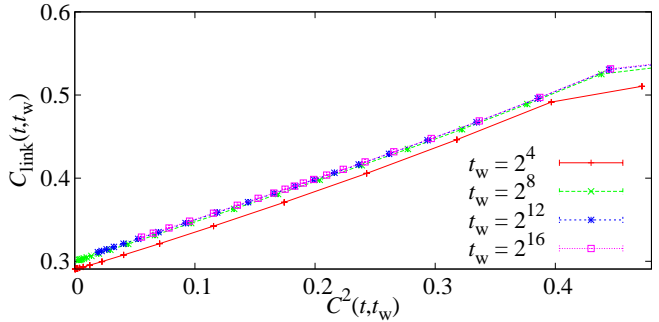}
\caption{$C_\mathrm{link}$ over $C^2(t,t_\mathrm{w})$ for different
  $t_\mathrm{w}$ at $T\!=\!0.7T_\mathrm{c}$ and for $D\!=\!22$.}\label{fig:Clink}
\end{center}
\end{figure}

Just as in the 3D case~\cite{PRLJANUS}, the aging dynamics in SG in the hypercube is a domain-growth process, see Fig \ref{fig:xi_70}. For any such process, the question of the ratio surface-volume arises. When this ratio vanishes in the limit of large domain size, as it is the case for any RSB dynamics, one expects a linear relation between $C_\mathrm{link}(t,t_\mathrm{w})$ and $C^2(t,t_\mathrm{w})$. This is precisely what we find in Fig. \ref{fig:Clink}.
\subsection{Thermoremanent magnetization}
\begin{figure}
\begin{center}
\includegraphics[angle=270,width=\columnwidth,trim=40 40 20 40]{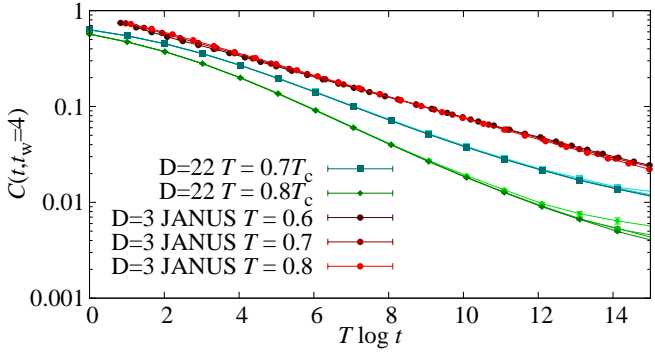}
\caption{Thermoremanent magnetization over $T\log t$. The JANUS data
  (in red circles), follow a power law with an exponent $\propto\!1/T$. Our
  results for $D\!=\!22$ are shown in dark tonalities (lighter colors: $D\!<\!22$).}\label{fig:termorremanente}
\end{center}
\end{figure}

The experimental work indicates that for
$T\!<\!0.9T_\mathrm{c}$, the thermoremanent magnetization follows a power law with an exponent proportional
to $T_\mathrm{c}/T$  \cite{THERMOREMANENT1}. The data obtained in JANUS for a
three dimensional SG (see Fig. \ref{fig:termorremanente} and \cite{JANUSJSP})
agree with this statement. However, the data obtained in the hypercube model
does not follow such power law, neither can them be rescaled with $T\log t$.

\begin{figure}
\begin{center}
\includegraphics[angle=270,width=\columnwidth,trim=40 40 20 40]{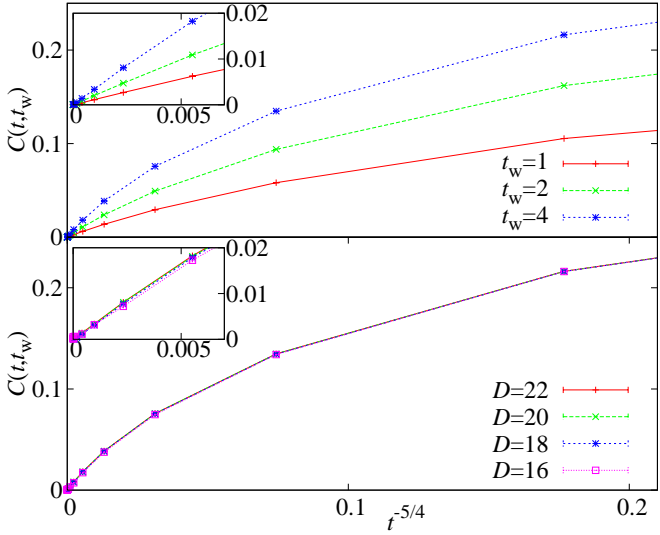}
\caption{Thermoremanent magnetization at $T_\mathrm{c}$ vs. $t^{-5/4}$, for
  \textbf{(up)} different $t_\mathrm{w}$ and $D=22$, and \textbf{(down)}
  different system sizes for $t_\mathrm{w}=4$. The two insets are close-ups of the origin.}\label{fig:termo}
\end{center}
\end{figure}

This lack of an algebraic decay is surprising on the view of the exact results of Ref. \cite{PRRR}. Indeed, it was analytically shown there that, at $T_\mathrm{c}$, the thermoremanent magnetization of the SK model decays as $t^{-5/4}$. Universality strongly suggests that the same scaling should hold for our model. Although it seems not to be the case, at the first glance, Fig. \ref{fig:termo}---top, a closer inspection confirms our expectation. Indeed, when plotted as a function of $t^{-5/4}$
, see inset in Fig. \ref{fig:termo}---top, the thermoremanent magnetization curve has a finite non-vanishing slope at the origin. As we show in bottom panel of Fig. \ref{fig:termo}, finite size effects do not contradict this claim. In summary, the magnetization decay for the hypercube suffers from quite strong finite time effects, but asymptotically it scales with the proper exponent, at least at $T_\mathrm{c}$.

\section{Nonequilibrium Correlation Functions and Finite Size Effects}\label{sec:TF}
\begin{figure}
\begin{center}
\includegraphics[angle=270,width=\columnwidth,trim=40 30 20 40]{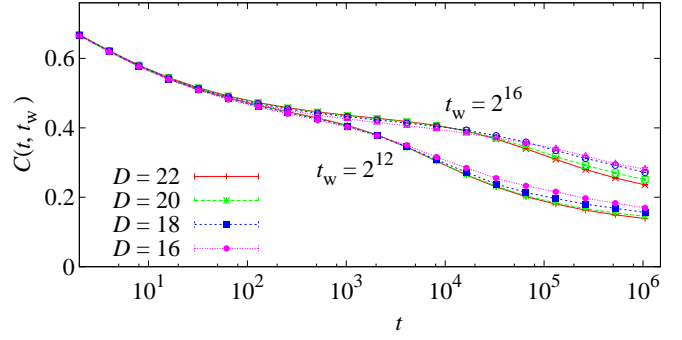}
\caption{Finite size effects in $C(t,t_\mathrm{w})$ at $T\!=\!0.7T_\mathrm{c}$.}\label{fig:Cttw}
\end{center}
\end{figure}

\begin{figure}
\begin{center}
\includegraphics[angle=270,width=\columnwidth,trim=10 30 20 40]{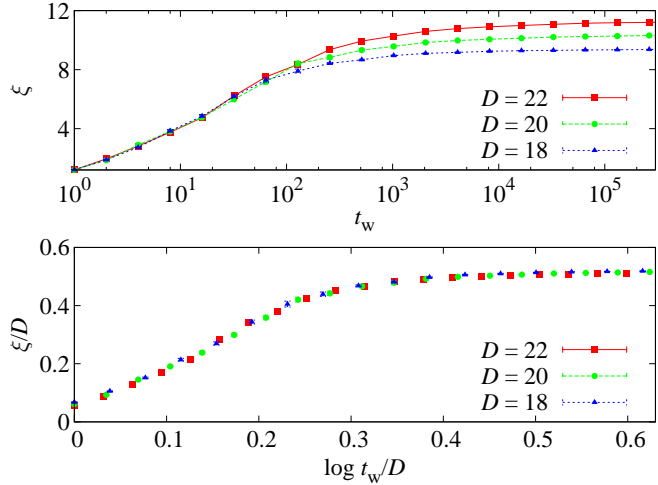}
\caption{\textbf{(Top)} Coherence length $\xi$ in the SG phase at $T=0.7T_\mathrm{c}$ vs. $t_\mathrm w$ for different system sizes. \textbf{(Bottom)} same data of the top panel rescaled by $D$ as a function of $\log t_\mathrm{w}/D$.}\label{fig:xi_70}
\end{center}
\end{figure}

The importance of finite size effects in nonequilibrium dynamics has been emphasized recently~\cite{PRLJANUS,JANUSJSP}. In our case, we have encountered important size effects, both in $C(t,t_\mathrm w)$, Fig. \ref{fig:Cttw}, and in $\xi(t_\mathrm w)$, Fig. \ref{fig:xi_70}--top.

\begin{figure}
\begin{center}
\includegraphics[angle=270,width=\columnwidth,trim=40 30 20 40
]{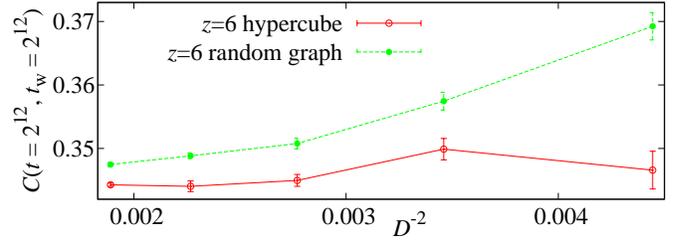}
\caption{$C(t,t_\mathrm{w})$ at $T\!=\!0.7T_\mathrm{c}$ for $t\!=\!t_\mathrm{w}\!=\!2^{12
}$ vs. $1/D^2$. We compare results obtained with two $z\!=\!6$ models: one
  with hypercubic topology \textbf{(red open circles)} and another in a totally random
  graph \textbf{(green full circles)}.}\label{fig:Cttw_kc_compara}
\end{center}
\end{figure}

We compare in Fig. \ref{fig:Cttw_kc_compara} the finite $D$ effects in $C(t,t_\mathrm w)$ for two different MF models with fixed connectivity: the hypercube and a previously studied model (the random graph with connectivity $z\!=\!6$, where each spin can interact with any other spin with
uniform probability~\cite{LEUZZI}). Clearly enough, the effects are much weaker in the hypercube model.

\begin{figure}
\begin{center}
\includegraphics[angle=270,width=\columnwidth,trim=40 30 20 40]{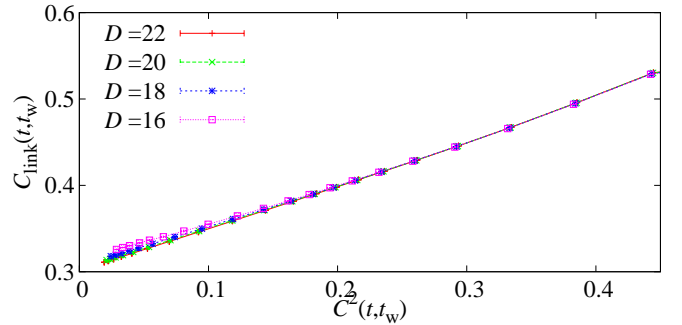}
\caption{$C_\mathrm{link}$ over $C^2(t,t_\mathrm{w})$ at
  $T\!=\!0.7T_\mathrm{c}$ for $t_\mathrm{w}=2^{12}$ and for different system sizes.}\label{fig:ClinkD}
\end{center}
\end{figure}

It is interesting to point out that, although the finite size effects
seems to be important in $C(t,t_\mathrm{w})$, they are largely absorbed when one eliminates the variable $t$ in favor of $C^2(t,t_\mathrm{w})$, see Fig. \ref{fig:ClinkD}. Hence, one of our main findings (the linear behavior of $C_\mathrm{link}$ as function of $C^2$) seems not endangered by finite size effects.

\begin{figure}
\begin{center}
\includegraphics[angle=270,scale=0.36,trim=70 30 20 40]{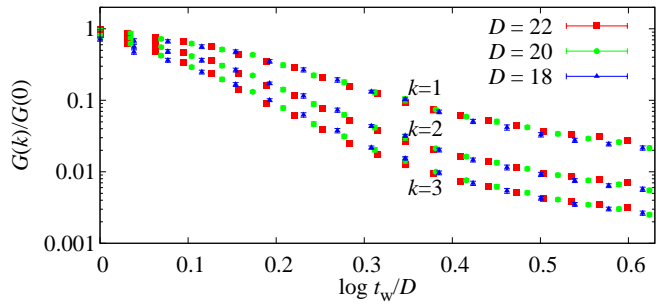}
\caption{Fourier transform $G(k)$ of $\hat{C}_4(r)$ in units of $G(0)$ as a function of $\log t_\mathrm{w}/D$ for several values of $D$ and $k$ at $T=0.7T_\mathrm{c}$. For each value of $k$, a different scaling function is found.}\label{fig:Gk_scaling}
\end{center}
\end{figure}

A very clear finite size effect is in the coherence length, $\xi(t_\mathrm{w})$. By definition, it cannot grow beyond $D$. Furthermore, what we find is that it hardly grows beyond $D/2$, Fig. \ref{fig:xi_70}--top. Nevertheless, at short times, we can identify a $D$-independent region, where it grows roughly as $\log t_\mathrm{w}$. Hence, one is tempted to conclude that $\xi_{D=\infty}(t_\mathrm{w})\propto\log t_\mathrm{w}$. At this point, finite size scaling suggests that both $\xi_D/D$ and $\log t_\mathrm{w}/D$ are dimensionless scaling variables. This is confirmed in Fig. \ref{fig:xi_70}--bottom, where a spectacular data collapse occurs. This is further confirmed by the Fourier transform of 
$\hat{C}_4(r)$, $G(k)$. Note that, since $\hat{C}_4(r)$ depends only on the length of the displacement vector $\V{r}$, also $G(k)$ is rotationally invariant. Now, since $k$ can range from 0 to $D$, it is clearly a dimensionless quantity (a dimesionful momentum would be $p=k/D$). It follows that $G(k)/G(0)$ is a dimensionless quantity that may depend only on a dimensionless variable, such as $\log t_\mathrm{w}/D$. Our data support this expectation, see Fig. \ref{fig:Gk_scaling}.

\begin{figure}
\begin{center}
\includegraphics[angle=270,scale=0.36,trim=40 30 20 40]{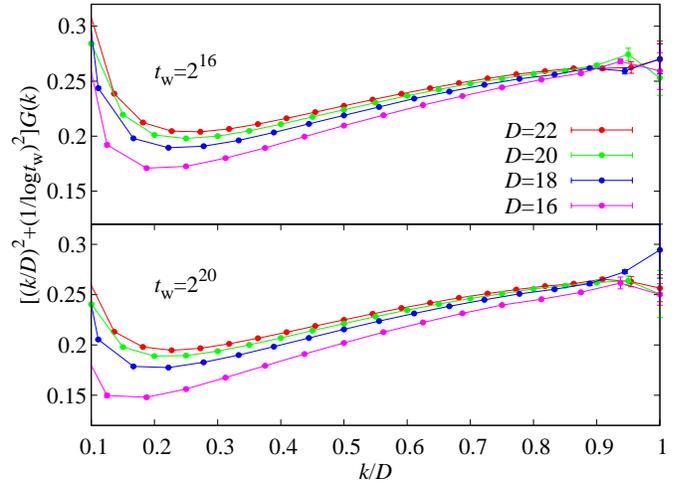}
\caption{Fourier transform $G(k)$ of $\hat{C}_4(r)$ in units of the
propagator [$\paren{p^2+1/\xi^2(t_\mathrm{w})}^{-1}$]~\cite{DEDOMINICIS} as a function of $p$,
where the dimensionful momentum is $p=k/D$ and $\xi(t_\mathrm{w})\sim \log
t_\mathrm{w}$. Recall that we are using postman metrics, hence, $p^2$
translates to $p^4$ in the Euclidean metrics. We show results for two waiting times: $t_\mathrm{w}=2^{16}$
\textbf{(top)} and $t_\mathrm{w}=2^{20}$
\textbf{(bottom)}. }\label{fig:Gk_scaling2}
\end{center}
\end{figure}

As for the $k$ dependence of $G(k)$, we expect a $1/p^4$
behavior in the range of $1/\xi(t_\mathrm{w})\ll p\ll
1$~\cite{DEDOMINICIS} (note that we are in the $q=0$ sector). Now, it is very important to recall that $p^4$ in
Euclidean metrics translates into $p^2$ in the postman metrics. We have also
seen that the dimensionful $p$ (postman metrics) corresponds to $k/D$. Thus,
since in our range of $t_\mathrm{w}$,
$\xi(t_\mathrm{w})\sim \log t_\mathrm{w}$, the product
$G(k)\paren{p^2+1/\log^2 t_\mathrm{w}}$ should be roughly constant as $D$ grows.
As we show in Fig. \ref{fig:Gk_scaling2}, the scaling is better for $p$ of
order 1 ($k\sim D$), although it seems to improve for smaller $p$ as $D$ grows.
As far as we know, this is the first observation of the $p^4$ propagator in a
numerical work.

\section{Conclusions}\label{sec:concl}
We have studied a spin glass model in the $D$-dimensional unit hypercube in the limit of large $D$, but with finite coordination number. We have shown that any short range model in such a lattice will behave as a mean field model in the thermodynamic limit (that coincides with the large $D$ limit). An important advantage of this model is that it has a natural notion of spatial distance.

We have shown that any statistical mechanics model on the hypercube with
random connectivity would be afflicted by huge finite size effects, for purely
geometrical reasons. The obvious cure has consisted in restricting the
connectivity graphs to those with a fixed number of neighbors. Unfortunately,
constructing such graphs is far from trivial. We have generated a subset of
them by means of a simple dynamic Monte Carlo. In this way, we obtain graphs
that are isotropic. We have checked that the Edwards-Anderson model defined over these finite connectivity hypercubes verify some consistency checks, including comparison with the analytically computable correlation function in the paramagnetic phase.

We have numerically studied the nonequilibrium dynamics in the spin glass
phase. The three main features found were: (i) aging dynamics consists in the
growth of a coherence length, much as in 3D systems, (ii) the scaling of the
two times correlation function implies infinitely many time-sectors, and (iii)
the $p^4$ propagator has been observed. In addition, we have studied the finite size effects in our model, finding that a naive finite size scaling ansatz accounts for our data.

\section*{Acknowledgments}
Computations have been carried out in PC clusters at BIFI and DFTI-UCM. We have been partly
supported through Research Contracts No. FIS2006-08533 (MICINN,
Spain) and UCM-BS, GR58/08. BS is an FPU fellow (Spain).

\appendix
\section{High temperature expansion}\label{ap:HTE}
For sake of clarity, we will firstly discuss the calculations for the random connectivity hypercube. Results for the fixed connectivity model will be then obtained by minor changes.

Using the identity ($\beta=1/T$)
\be e^{\beta J_{\V{x}\V{y}}\sigma_{\V{x}}\sigma_{\V{y}}}=\cosh{\beta}\paren{1+J_{\V{x}\V{y}}\sigma_{\V{x}}\sigma_{\V{y}}\tanh{\beta}},\ee
we can write the partition function and the spin propagator as:
\begin{eqnarray}
 \frac{Z}{(\cosh{\beta})^{DN}}&=&\sum_{\lazo{\sigma}}\prod_{\mean{\V{z}\V{w}}}\paren{1+J_{\V{z}\V{w}}\sigma_{\V{z}}\sigma_{\V{w}}\tanh{\beta}},\\
\sigma_{\V{x}}\sigma_{\V{y}}&=&\\
&&\frac{\sum_{\lazo{\sigma}}\sigma_{\V{x}}\sigma_{\V{y}}\prod_{\mean{\V{z}\V{w}}}\paren{1+J_{\V{z}\V{w}}\sigma_{\V{z}}\sigma_{\V{w}}\tanh{\beta}}}{\sum_{\lazo{\sigma}}\prod_{\mean{\V{z}\V{w}}}\paren{1+J_{\V{z}\V{w}}\sigma_{\V{z}}\sigma_{\V{w}}\tanh{\beta}}}.\nonumber
\end{eqnarray}
The high-temperature expansion (see, for instance~\cite{PARISI-SFT}), expresses the propagator as a sum over lattice paths that join the points $\V{x}$ and $\V{y}$, $\gamma_{\V{x}\rightarrow\V{y}}$:
\be\mean{\sigma_{\V{x}}\sigma_{\V{y}}}=Z^{-1}\sum_{\gamma_{\V{x}\rightarrow\V{y}}}Z_\gamma J(\tanh\beta)^{l_\gamma},\ee
where $l_\gamma$ represents the length of the path
$\gamma_{\V{x}\rightarrow\V{y}}$, $J$ is the product of the couplings,
$J_{\V{z}\V{w}}$, along the path, and $Z_\gamma$ is a restricted partition
function obtained by summing only over all closed paths that do not have any
common link with the path $\gamma_{\V{x}\rightarrow\V{y}}$.

However, when averaging over disorder, due to the randomness in the coupling signs,
$\overline{\mean{\sigma_{\V{x}}\sigma_{\V{y}}}}=0$.
The spin glass propagator is obtained instead by averaging over disorder $\mean{\sigma_{\V{x}}\sigma_{\V{y}}}^2$. Clearly, the sum will be dominated by those diagrams where the go and return path are the same (thus, $J_{\V{z}\V{w}}^2=1$):
\be\overline{\mean{\sigma_{\V{x}}\sigma_{\V{y}}}^2}=Z^{-2}\sum_{\gamma_{\V{x}\rightarrow\V{y}}}
Z_\gamma^2\caja{\tanh^2{\beta}}^{l_\gamma}=Z^{-2}\sum_{\gamma_{\V{x}\rightarrow\V{y}}}K^{l_\gamma}\, Z_\gamma^2,\ee
where $K=\tanh^2{\beta}$. In Bethe lattices, due to their cycle-less nature, $Z_\gamma^2/Z^2=1$ in the thermodynamic limit. Hence, we are left with the problem of counting the average number of paths of length $l_\gamma$ that join $\V{x}$ and $\V{y}$, $p(l_\gamma)$. From it, we obtain 
\be\label{eq:C4Ap} \hat{C}_4(r)=\binom{D}{r}\sum_{l_\gamma\ge r}p(l_\gamma)K^{l_\gamma}.\ee
The sum is restricted to $l_\gamma\ge r$ because the length of the shortest path that joins $\V{x}$ and $\V{y}$ is given by their postman distance $r$.

In order to count the average number of paths, $p(l_\gamma)$, let us distinguish two cases: $l_\gamma=r$ and $l_\gamma>r$. The first will give the leading contribution in the large $D$ limit.

 The number of paths joining $\V{x}$ and $\V{y}$ in precisely $r$ steps is $r!$, because the $r$ steps are all taken along different directions and in a random order. For a given path, the probability of all the $r$ links be active is $(z/D)^r$. Hence
\be p\paren{l_\gamma=r}=\frac{z^r}{D^r}r!\,.\ee
Note that the $D^{-r}$ factor compensates exactly the divergence of the $\binom{D}{r}$ in Eq. \eqref{eq:C4Ap}.

In the case of $l_\gamma>r$, one has $l_\gamma=r+2k$, with $k>0$. Note that when $l_\gamma=r$ the path contains $r$ different directions (namely, the Euclidean components in which $\V{x}$ and $\V{y}$ differ). Each of these directions appear only once. However, when $l_\gamma>r$, other directions must be included, we call them unnecessary. Note that, if the path is to end at the desired point, any unnecessary step must be undone later on. Hence, $l_\gamma-r$ is always an even number $2k$. Clearly, the number of such paths is bounded by $\Gamma\paren{r,k}D^k$, where $\Gamma\paren{r,k}$ is a $D$-independent amplitude. On the other hand, the probability of finding all the links active is $(z/D)^{r+2k}$. Thus, we conclude that 
\be p\paren{l_\gamma=r+2k}=O\paren{\frac{1}{D^{k+r}}},\ee
that results in a $O\paren{D^{-k}}$ contribution to $\hat{C}_4 (r)$.

Then, in the large $D$ limit we obtain ($A=zK$):
\be \hat{C}_4(r)=A^r=e^{r \log A},\ee
with finite size corrections of $O\paren{D^{-1}}$.
Thus, we encounter an exponential decay with an exponential correlation length given by
\be\xi^\text{exp}=\frac{1}{|\log A|}.\ee

Summing all up, we can compute the spin-glass susceptibility for the large $D$ limit:
\be\chi=\sum_{r=0}^\infty\hat{C}_4(r)=\sum_{r=0}^\infty A^r=\frac{1}{1-A}.\ee
We see that when $A=1$ the correlation no longer decays with distance, and the susceptibility diverges. Of course, one gets $A=1$ precisely at the critical temperature, $T_\mathrm c$, reported in Eq. \eqref{eq:KC}.

The computation for the fixed connectivity model is 
very similar. One only needs to notice that, whereas the probability for the first link in a lattice path to be active is $z/D$, the probability for the next link is roughly $(z-1)/D$ (this is only accurate for large $D$). It follows that, again, the $l_\gamma=r$ paths are the only relevant paths in the high temperature expansion. We find that 
\be p(l_\gamma=r)=\left\{\begin{array}{ll}
 1&\text{ if  }\ r=0,\\
\frac{z}{D}\paren{\frac{z-1}{D}}^{r-1}r!&\text{ if  }\ r>0.
\end{array}\right.\ee
Again, we can use it to compute $ \hat{C}_4(r)$. In the
large $D$ limit, up to corrections of $O\paren{D^{-1}}$, it is given by:
\be
\hat{C}_4(r)=\left\{\begin{array}{lc}1&\text{ if  }\  r=0,\\\frac{z}{z-1}\caja{\paren{z-1}K}^r&\text{ if  }\ r>0,\end{array}\right.\ee
which, taking $\tilde A=(z-1)K$, also shows an exponential decay with 
\be\xi^\text{exp}=\frac{1}{|\log \tilde A|}.\ee

Using this spatial correlation function, we can either compute the SG-susceptibility in the fixed connectivity hypercube,
\be\label{eq:chi_inftyAP}\chi=\sum_{r=0}^\infty\hat{C}_4(r)=1+\frac{z}{z-1}\frac{\tilde A}{1-\tilde A},\ee
or the integral correlation length, defined as \eqref{eq:xi},
\be\label{eq:xi_infty}\xi=\frac{\sum_{r=0}^\infty r\,\hat{C}_4(r)}{\sum_{r=0}^\infty\hat{C}_4(r)}=\frac{\chi-1}{\chi}\frac{1}{1-\tilde A}.\ee

Again, when $\tilde A=1$. we find a critical point. The corresponding $T_\mathrm c$ matches  Eq. \eqref{eq:KC}. The critical exponents, $\gamma=1$, $\nu=1$, can be read directly from Eq. \eqref{eq:chi_inftyAP} and \eqref{eq:xi_infty}. The reader might be puzzled by a mean field model with $\nu\neq 1/2$. The solution to the paradox is in our chosen metrics. Recall that the postman distance in the hypercube is the square of the Euclidean one. Hence, the correlation length in Eq. \eqref{eq:xi_infty} is the \textit{square} of the Euclidean correlation length.

\section{Scaling and dynamic ultrametricity}\label{ap:Ultrametricity}
\begin{figure}
\begin{center}
\includegraphics[angle=270,width=\columnwidth,trim=80 30 30
  22]{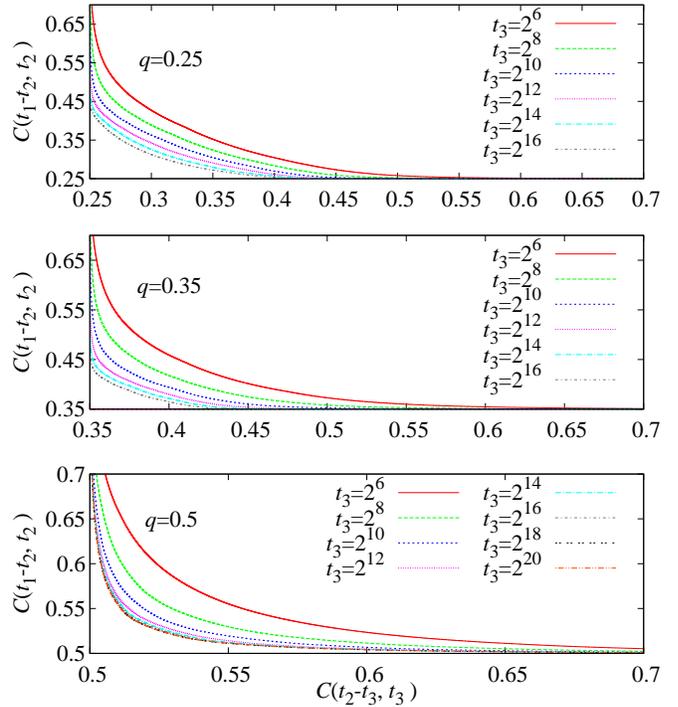}
\caption{Parametric plot
  $\caja{x(t_2),y(t_2)}=\caja{C(t_1-t_2,t_2),C(t_2-t_3,t_3)}$, $t_1>t_2>t_3$
  with $t_1$ fixed by the condition $C(t_1-t_3,t_3)=q$ and different
  $t_3$. In the presence of dynamic ultrametricity, \eqref{eq:cond}, the parametric plot should
  tend for large $t_3$ to the union of $x=q$ and $y=q$.
The panels correspond to $q=0.25$ (\textbf{top}, nice BB scaling but not
  ultrametric), $q=0.35$ (\textbf{middle}, nice BB scaling and ultrametric) and $q=0.5$
(\textbf{bottom}, supposedly ultrametric but poor BB scaling).}\label{fig:C12vsC23}
\end{center}
\end{figure}
\begin{figure}
\begin{center}
\includegraphics[angle=270,width=\columnwidth,trim=50 30 30
  22]{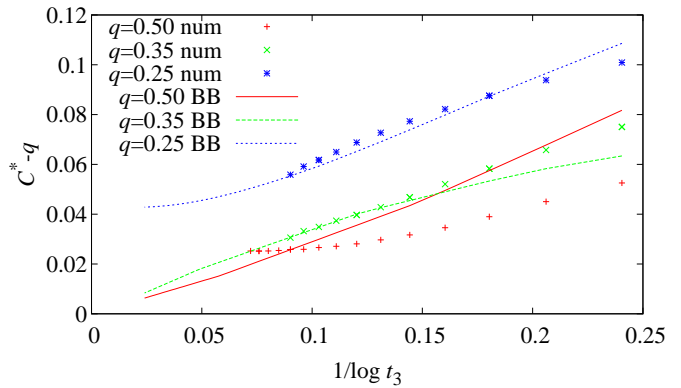}
\caption{
\textbf{Dots:} For each $q$ and $t_3$, as in Fig. \ref{fig:C12vsC23},
  we take the intercept with $x=y$, i.e. $C^*=C(t_1-t_2,t_2)=C(t_2-t_3,t_3)$,
  and represent $C^*-q$ as a function of $1/\log t_3$. \textbf{Lines:}
   analogous plot for the toy model described in the text,
  where the BB scaling is exact. }\label{fig:C12vst3}
\end{center}
\end{figure}
\begin{figure}
\begin{center}
\includegraphics[angle=270,width=\columnwidth,trim=20 30 30 22]{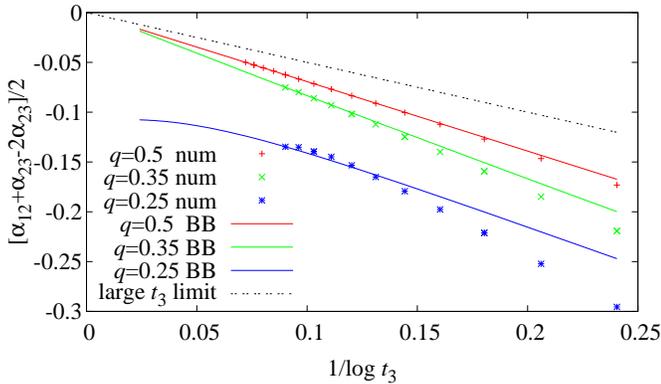}
\caption{For the data in Fig. \ref{fig:C12vst3}, we represent
  $\frac{\alpha(t_1,t_2)+\alpha(t_2,t_3)}{2}-\alpha(t_1,t_3)$ vs. $1/\log
  t_3$. The dashed line corresponds to Eq. \eqref{eq:alpha}.   }\label{fig:alpha12}
\end{center}
\end{figure}
As in Eq. \eqref{eq:ultrametric1}, let us assume that the spin time correlation function behaves for large $t_{\mathrm w}$ as 
\be C(t,t_\mathrm{w})=f\paren{\alpha(t,t_\mathrm{w})}\,,\quad \alpha(t,t_\mathrm{w})=\log t/\log t_\mathrm{w}\,,\ee
where the scaling function $f$ is smooth and monotonically decreasing. From
now on, we shall refer to this scaling as BB scaling (after Bertin-Bouchaud).

Let us see under which conditions BB scaling implies the ultrametricity property
\be\label{eq:cond} C(t_1-t_3,t_3)= \min\left\{C(t_1-t_2,t_2),C(t_2-t_3,t_3)\right\},\ee
where $t_1\gg t_2\gg t_3$ and $t_3$ tends to infinity.

The natural time dependency is a power law choice
\begin{eqnarray}
t_1&=&t_3+At_3^{\mu_1},\\
t_2&=&t_3+Bt_3^{\mu_2},
\end{eqnarray}
with $\mu_1>\mu_2$. In that case, the large $t_3$ limit for the argument of
the scaling function are: $\alpha(t_1-t_3,t_3)=\mu_1$,
$\alpha(t_2-t_3,t_3)=\mu_2$ and $\alpha(t_1-t_2,t_2)=\mu_1$ if $\mu_2 < 1$ and
$\alpha(t_1-t_2,t_2)=\mu_1/\mu_2$ if $\mu_2 > 1$. Then, the condition
\eqref{eq:cond} is only satisfied in case $\mu_2 < 1$. If, as it is the case
for the critical trap model~\cite{BOUCHAUD},
$f(\alpha>1)=\text{constant}$~\footnote{Weak ultrametricity breaking implies
  that $f(\alpha>1)=0$.}, the BB scaling would imply dynamic
ultrametricity. This is not the case for a general scaling function $f$ such
as, for instance, the one we get in Fig. \ref{fig:ultrametric1}.
Nevertheless, although this analysis implies that the dynamic ultrametricity
is only present in our model in some range of parameters, let us try a more
straight approach. 

We consider a fixed value for the correlation function, $q$. On the view of
the previous considerations and of Fig. \ref{fig:ultrametric1}, we should
expect ultrametricity only for $q>f(\alpha=1)\approx0.35$. Now, for each
$t_3$, we find $t_1$ such that $C(t_1-t_3,t_3)=q$. Then, we perform a
parametric plot of $C(t_1-t_2,t_2)$ vs. $C(t_2-t_3,t_3)$, for
$t_3<t_2<t_1$. Ultrametricity predicts that, in the large $t_3$ limit, the
curves should tend to a half square (e.g. the intersection of the straight
lines $x=q$ and $y=q$) and, in particular, when $C(t_1-t_2,t_2)=C(t_2-t_3,t_3)=C^*$, $C^*$ should tend to
$q$. 

We present in Fig. \ref{fig:C12vsC23} results for three different values
of $q$: 0.5 (ultrametric region, but in our range of $t_\mathrm{w}$ data do
not scale according BB), 0.35 (ultrametric region and good BB scaling) and
0.25 (non ultrametric region but BB scaling works nicely). At the qualitative
level, the parametric curves seem to tend to a corner, but the convergence is
slow. Furthermore, there are no clear differences between the curves with
$q>f(\alpha=1)$ and those with $q<f(\alpha=1)$. Hence, we may try a more quantitative analysis.

We obtain numerically $C^*$, the point where
$C^*=C(t_1-t_2,t_2)=C(t_2-t_3,t_3)$, and study $C^*-q$ as function of $1/\log
t_3$ in Fig. \ref{fig:C12vst3}. This choice is due to the fact that in the
ultrametric region BB scaling predicts
\be\label{eq:alpha}\alpha(t_1-t_2,t_2)=\alpha(t_1-t_3,t_3)+\frac{1} {2\log t_3} +\ldots\,.\ee Hence,
we expect that $C^*-q$ will be of order $1/\log t_3$ if ultrametricity
holds. The numerical data confirms this picture only partly. For $q=0.35$ the
results are as expected, yet for $q=0.25$ the difference is decreasing fast as
$t_3$ grows and it is hard to tell whether the extrapolation will be zero or
not. For $q=0.5$ (where BB scaling is not working for our numerical data) the
behavior is non monotonic.

To rationalize our finding, we consider a
simplified model, where the BB scaling is supposed to hold exactly. The
master curve $f(\alpha)$ is taken from the numerical data for
$C(t,t_3=2^{16})$ for $D=22$. This toy model allows us consider ridiculously
large values of $t_3$. As we see in Fig. \ref{fig:C12vst3}, the peculiarities
of the master curve cause a non monotonic behavior in $q$ for an ample range
of $t_3$.

The lack of monotonicity in $q$ makes also on interest to focus on $\alpha$,
rather than on the correlation function. With this aim, we consider the time
$t_2$ where $C(t_1-t_2,t_2)=C(t_2-t_3,t_3)=C^*$, and compute
$\frac12\caja{\alpha(t_1-t_2,t_2)+\alpha(t_2-t_3,t_3)}-\alpha(t_1-t_3,t_3)$. BB
scaling and ultrametricity combined, see Eq. \eqref{eq:alpha}, imply that this
quantity should be of order $1/\log t_3$ (in the non ultrametric region, it
should be of order one).  Our results in Fig. \ref{fig:alpha12} basically
agree with these expectations.

\end{document}